\documentclass[journal]{IEEEtran}

\usepackage{amssymb}
\usepackage[cmex10]{amsmath}
\usepackage{stfloats}
\usepackage{graphicx}
\usepackage{subfigure}
\usepackage{tabularx}
\usepackage{epsfig,epsf,color,balance,cite}
\usepackage{verbatim}
\usepackage{url}
\usepackage{bm}

\newtheorem{theorem}{Theorem}

\usepackage{algorithm}

\usepackage{algpseudocode}
\usepackage{color}
\definecolor{myc1}{rgb}{0,0,0}
\definecolor{myc2}{rgb}{0,0,0}
\begin{document}

\title{Compressive Sensing Based User Clustering for Downlink NOMA Systems with Decoding Power}
\author{
\IEEEauthorblockN{Zhaohui Yang,
                  Cunhua Pan,
                  Wei Xu, \IEEEmembership{Senior Member, IEEE},
                   and
                  Ming Chen
                  }
                  \vspace{-3em}
\thanks{Z. Yang, W. Xu and M. Chen are with the National Mobile Communications Research
Laboratory, Southeast University, Nanjing 210096, China  (Email: \{yangzhaohui, wxu, chenming\}@seu.edu.cn).}
 \thanks{C. Pan is with the School of Electronic Engineering and Computer Science, Queen Mary, University of London, London E1 4NS, U.K. (Email: c.pan@qmul.ac.uk).}
}
\maketitle

\begin{abstract}
This letter investigates joint power control and user clustering for downlink non-orthogonal multiple access systems.
Our aim is to minimize the total power consumption by taking into account not only the conventional transmission power but also the decoding power of the users.
To solve this optimization problem, it is firstly transformed into an equivalent problem with tractable constraints.
Then, an efficient algorithm is proposed to tackle the equivalent problem by using the techniques of {\color{myc2}{reweighted}} $\ell_1$-norm minimization and majorization-minimization.
Numerical results validate the superiority of the proposed algorithm over the conventional algorithms including the popular matching-based algorithm.
\end{abstract}
\begin{IEEEkeywords}
Non-orthogonal multiple access, power control, user clustering, majorization-minimization.
\end{IEEEkeywords}

\IEEEpeerreviewmaketitle

\section{Introduction}
\label{section1}

Non-orthogonal multiple access (NOMA) has been deemed as a promising technology for future fifth generation systems \cite{Dai2015Non,ding2017application,6666209}.
By applying superposition coding at the transmitter and employing successive interference cancellation (SIC) at the users, NOMA serves multiple users with the same time-frequency resource, which makes NOMA more spectral efficient than orthogonal multiple access (OMA).
In \cite{ding2014performance}, it was shown that NOMA can achieve superior performance in terms of ergodic sum rate compared with OMA.
The power control problem of maximizing sum rate was investigated in \cite{Yang2017On} for downlink NOMA systems.
Besides, the impact of user pairing in NOMA with fixed power allocation was investigated in \cite{Ding2016Impact}, which showed that the paired users should have distinctive channel gains to obtain large sum rate.
Moreover, the authors in \cite{Fang2016EENOMA} studied joint power control and user pairing, where the user pairing subproblem was solved by using matching theory.

In addition to spectral efficiency, power minimization for NOMA has been attracting research attention lately.
To minimize total transmission power, a distributed power control algorithm was proposed by using the game theoretic approach \cite{Sung2016Game} for uplink NOMA.
In \cite{Fu2016Distributed}, the standard interference function was applied to solve the downlink sum power minimization problem for a two-cell NOMA system.
It is proven in \cite{Lei2016OnPower} that the downlink power minimization problem for NOMA with multiple subcarriers is NP-hard, and a relax-then-adjust algorithm was accordingly proposed.
However, the above works \cite{Sung2016Game,Fu2016Distributed,Lei2016OnPower} all ignored the decoding power in the power consumption model, even though the decoding power is comparable to the transmission power \cite{Xiong2012Decoding}.

In this letter, we investigate the downlink power minimization problem for NOMA.
{\color{myc2}{
There are two main contributions in this letter.
One contribution is that we consider the decoding power consumption, which is important but ignored in many existing works.
The other contribution is to tackle this nonconvex power minimization problem. In particular, we first successfully transform the original problem with nonlinear rate constraints into an equivalent problem with linear rate constraints, and then adopt the penalty method and compressive sensing method to solve a sequence of tractable convex problems.
}}

\section{System Model and Problem Formulation}
\label{section2}
\subsection{System Model}
Consider a downlink NOMA system with one single-antenna base station (BS) and $M$ single-antenna users. 
In this system, there are $N$ subcarriers.
The channel gain between the BS and the $m$-th user on the $n$-th subcarrier is denoted by $h_{mn}$.
Without loss of generality, the channels are sorted as $|h_{1n}|^2\leq|h_{2n}|^2\leq\cdots\leq|h_{Mn}|^2$, for all $n=1, \cdots, N$.
To ensure the sort of channel gains, user $m$ is denoted as the $m_n$-th user on the $n$-th subcarrier.


According to the NOMA principle, the BS simultaneously transmits signal to all the users.
The transmitted signal $x_n$ on the $n$-th subcarrier can be expressed as
\begin{equation}
\vspace{-0.25em}
x_n=\sum_{m=1}^M\sqrt{p_{mn}}x_{mn},
\vspace{-0.25em}
\end{equation}
where $x_{mn}$ and $p_{mn}$ are the message and allocated power for the $m$-th user on the $n$-th subcarrier, respectively.
The observation at the $m$-th user on the $n$-th subcarrier is
\begin{equation}\label{eq1}
y_{mn}= h_{mn}\sum_{l=1}^M\sqrt{p_{ln}} x_{ln}+z_{mn},
\end{equation}
where $z_{mn}$ represents the additive zero-mean Gaussian noise with variance $\sigma^2$.

{\color{myc1}{
For downlink NOMA, SIC is carried out at the users.
Assume that the bandwidth for each subcarrier is $B$.
According to \cite{ding2014performance}, the achievable rate of the $m$-th user on the $n$-th subcarrier is
\begin{equation}\label{eq2}
r_{mn}=B\log_2\left(
1+\frac{H_{mn} p_{mn}}
{H_{mn} \sum_{l=m+1}^{M} p_{ln} +\sigma^2}
\right)
\end{equation}
for $1\leq m \leq M-1$, and
\begin{equation}\label{eq2_1}
r_{Mn}=B\log_2\left(
1+\frac{H_{Mn} p_{Mn}}
{\sigma^2}
\right),
\end{equation}
where $H_{mn}=|h_{mn}|^2$, $\forall m =1,\cdots, M$.
}}

Since one user can be allocated with multiple subcarriers, the achievable rate of user $m$ is given by
\begin{equation}\label{eq3}
r_{m}=\sum_{n=1}^N r_{m_n n}.
\end{equation}

\subsection{Power Consumption Model}
The total power consumption of the system consists of two parts: the transmission power of the BS and the decoding power of the users.
By summing the transmission power for all users on all subcarriers, the total transmission power of the BS can be calculated as 
\begin{equation}\label{sysPowerBSl}
\sum_{m=1}^M\sum_{n=1}^N p_{mn}.
\end{equation}

{\color{myc1}{
According to \cite{Rubio2014Energy}, the decoding complexity of each user grows linearly with the decoded rate of each user.
A linear function relating to the rate and the power consumed by the decoder at user $m$ can be given by \cite{Rubio2014Energy,Nguyen2016EE,VLSI2015CGTIT}
\begin{equation}\label{REv2dec1}
P_m^{\text{dec}}(\bar r_m)=\lambda_m \bar r_m,
\end{equation}
where $\lambda_m$ is the decoder efficiency of user $m$,
and $\bar r_m$ is the decoded rate of user $m$.
Since user $m$ should decode the signals of weak users before decoding its own message \cite{ding2014performance}, the decoded rate of user $m$ on the $n$-th subcarrier is $\sum_{s=1}^{m_n} r_{sn}$ if user $m$ occupies the $n$-th subcarrier, i.e., $p_{m_nn}>0$.
According to (\ref{REv2dec1}), the decoding power of user $m$ is
\begin{equation}\label{REv2dec2}
P_m^{\text{dec}}(\bar r_m)=\sum_{n=1}^N \sum_{s=1}^{m_n} \lambda_m r_{sn}\|p_{m_nn}\|_0,
\end{equation}
where $\|\cdot\|_{0}$ is the $\ell_0$-norm.}}

The total power consumption, denoted as $P_{\text{total}}$, is given by
\begin{equation}\label{sysPtotal}
P_{\text{total}}=\sum_{n=1}^N \sum_{m=1}^M p_{mn}+
{\color{myc1}{ \sum_{n=1}^N \sum_{m=1}^M \sum_{s=1}^{m_n} \lambda_m r_{sn}\|p_{m_nn}\|_0}}.
\end{equation}
From (\ref{sysPtotal}), the total power consumption contains two parts that conflict each other.
On one hand, the transmission power part decreases with the number of users occupying one subcarrier (according to \cite{Lei2016OnPower}).
{\color{myc1}{This is due to the fact that users can occupy more subcarriers and SIC is helpful in mitigating inter-user interference for larger number of users occupying one subcarrier.}}
On the other hand, the decoding power part increases with the number of users occupying one subcarrier for higher decoded rate of the users.

\subsection{Problem Formulation}
According to (\ref{eq2}), (\ref{eq2_1}), (\ref{eq3}) and (\ref{sysPtotal}), the total power optimization problem can be formulated as:
\begin{subequations}\label{max1_1}
\begin{align}
\mathop{\min}_{ \pmb{p}\geq\pmb{0}  }& \quad
\sum_{n=1}^N\sum_{m=1}^M p_{mn}+
{\color{myc1}{ \sum_{n=1}^N \sum_{m=1}^M \sum_{s=1}^{m_n} \lambda_m r_{sn}\|p_{m_nn}\|_0}}
\\
\textrm{s.t.}
&\quad \sum_{n=1}^N r_{m_n n} = R_m, \quad\forall m\\
&\quad \|\pmb p_n\|_{0} \leq L, \quad \forall n,
\end{align}
\end{subequations}
where $\pmb p=[p_{11}, \cdots, p_{M1}, \cdots, p_{MN}]^T$,
$r_{sn}$ is the achievable rate of the $s$-th user on the $n$-th subcarrier defined in (\ref{eq2}),
$R_m$ is the rate demand of user $m$,
$L$ is the maximum number of multiplexed users on each subcarrier,
and $\pmb p_n=[p_{1n}, \cdots, p_{Mn}]^T$.
Due to the practical limitations of the receiver's design complexity
and the signal processing time for SIC, 
$L$ is a parameter with $L\leq M$.

\section{Joint Power Allocation and User Clustering}

\subsection{Equivalent Transformation}
%

\begin{theorem}
The original total power minimization Problem (\ref{max1_1}) can be equivalently transformed to the following problem as:
\begin{subequations}\label{max3_1}
\begin{align}
\mathop{\min}_{ \pmb{r}\geq\pmb 0}& \:\;
\sum_{n=1}^N\sum_{m=1}^{M}
\left(\frac{ \sigma^2}
{H_{mn} }-\frac{ \sigma^2}
{H_{(m+1)n} }\right){2^{\sum_{s=1}^{m}\frac{r_{sn}}{B}}}
\nonumber\\
& +   {\color{myc1}{\sum_{n=1}^N \sum_{m=1}^M \sum_{s=1}^{m_n} \lambda_m r_{sn}\|r_{m_nn}\|_0}}
\\
\textrm{s.t.}
& \:\;\sum_{n=1}^N r_{m_n n} = R_m, \quad\forall m\\
& \:\; \|\pmb r_n\|_{0} \leq L, \quad \forall n,
\end{align}
\end{subequations}
\end{theorem}
where $\pmb r\!=\![r_{11}, \cdots,$ $ r_{M1}, \cdots, r_{MN}]^T$, $\pmb r_n=[r_{1n}, \cdots, r_{Mn}]^T$, and we set $\frac{\sigma^2}{H_{(M+1)n}}=0$ for all $n$.

\itshape \textbf{Proof:}  \upshape
Please refer to Appendix A.
 \hfill $\Box$

From Theorem 1, nonlinear constraints (\ref{max1_1}b) in Problem (\ref{max1_1}) are converted to linear constraints (\ref{max3_1}b) in Problem (\ref{max3_1}).
To deal with constraints (\ref{max3_1}c) with non-smooth $\ell_0$-norm, we adopt the penalty method \cite{boyd2004convex}.
Specially, Problem (\ref{max3_1}) can be reformulated as:
\begin{subequations}\label{max3_1_1}
\begin{align}
\mathop{\min}_{ \pmb{r}\geq\pmb 0}& \: \sum_{n=1}^N\sum_{m=1}^{M}
\left(\frac{ \sigma^2}
{H_{mn} }-\frac{ \sigma^2}
{H_{(m+1)n} }\right){2^{\sum_{s=1}^{m}\frac{r_{sn}}{B}}}
\nonumber\\
&
\!+\!
 {\color{myc1}{\sum_{n=1}^N \sum_{m=1}^M \sum_{s=1}^{m_n} \lambda_m r_{sn}\|r_{m_nn}\|_0}}
+\sum_{n=1}^N\!\frac{\|\pmb r_n\|_0^K} {(L+0.5)^K}
\\
\textrm{s.t.}
&\:\;\sum_{n=1}^N r_{m_n n} = R_m, \quad\forall m,
\end{align}
\end{subequations}
where $K$ is a large positive constant.
When $K$ becomes infinity, Problem (\ref{max3_1}) is equivalent to Problem (\ref{max3_1_1}) with fewer constraints.
The reason is that $\lim_{K \rightarrow +\infty}\left(\frac{\|\pmb r_n\|_0}{L+0.5}\right)^K=0$ for $\|\pmb r_n\|_0\leq L$, and  $\lim_{K \rightarrow +\infty}\left(\frac{\|\pmb r_n\|_0}{L+0.5}\right)^K=+\infty$ for $\|\pmb r_n\|_0>L$.
\subsection{{\color {myc1}{Compressive Sensing}} Based Algorithm}
The difficulty to solve Problem (\ref{max3_1_1}) is the non-smooth $\ell_0$-norm in the objective function (\ref{max3_1_1}a).
%
%
%
{\color{myc1}{Since $\|\pmb r_n\|\leq L$ and $L$ is usually small in practical situations, $\pmb r_n$ can be viewed as a sparse vector in the compressive sensing {\color{myc2}{method}}. To deal with (\ref{max3_1_1}a), non-smooth $\ell_0$-norm minimization can be approximately solved via a sequence
of weighted $\ell_1$-norm minimizations in compressive sensing according to \cite{candes2008enhancing} and \cite{7437385}.}}
{\color{myc2}{Specifically, non-smooth $\ell_0$-norm is approximated by
\vspace{-0.25em}
\begin{equation}\label{appenBeq1}
\|x\|_0=\lim_{\tau\rightarrow 0}\frac{\ln(1+x\tau^{-1})}{\ln(1+\tau^{-1})}, \quad \forall x\geq 0.
\end{equation}
\vspace{-0.25em}
Replacing non-smooth $\ell_0$-norm in Problem (\ref{max3_1_1})  with the logarithmic function according to (\ref{appenBeq1}), we have
\vspace{-0.25em}
\begin{subequations}\label{max3_1appenB}
\begin{align}
\mathop{\min}_{ \pmb{r}\geq\pmb 0}& \:\;
\sum_{n=1}^N\sum_{m=1}^{M}
\left(\frac{ \sigma^2}
{H_{mn} }-\frac{ \sigma^2}
{H_{(m+1)n} }\right){2^{\sum_{s=1}^{m}\frac{r_{sn}}{B}}}
 + \sum_{n=1}^N \sum_{m=1}^M \sum_{s=1}^{m_n} \nonumber\\
&\lambda_m r_{sn}\frac{\ln(1+r_{m_nn}\tau^{-1})}{\ln(1+\tau^{-1})}
+\sum_{n=1}^N\frac{\left(\sum\limits_{m=1}^M\ln(1+ r_{mn}\tau^{-1})\right)^K}{(L+0.5)^K\ln^K(1+\tau^{-1})}
\\
\textrm{s.t.}
& \:\;\sum_{n=1}^N r_{m_n n} = R_m, \quad\forall m.
\end{align}
\end{subequations}
\vspace{-0.25em}


Using (\ref{appenBeq1}) and the first-order approximation,
we approximate the $\ell_0$-norm  in the objective function (\ref{max3_1_1}a) as
\begin{equation}\label{paeq1}
\|\pmb r_n\|_{0}\approx
\sum_{m=1}^M (w_{mn}^{(t)} r_{mn}+\alpha_{mn}^{(t)}),
\end{equation}
with $w_{mn}^{(t)}$ and $\alpha_{mn}^{(t)}$ iteratively updated according to
\begin{equation}\label{paeq2}
w_{mn}^{(t)}=\frac{1}{(r_{mn}^{(t)}+\tau)\ln(1+\tau^{-1})},
\end{equation}
and
\begin{equation}\label{paeq2_1}
\alpha_{mn}^{(t)}=\frac{(r_{mn}^{(t)}+\tau)\ln(1+\tau^{-1}r_{mn}^{(t)})-r_{mn}^{(t)}}
{(r_{mn}^{(t)}+\tau)\ln(1+\tau^{-1})},
\end{equation}}}
\!\!\!where $r_{mn}^{(t)}$ is value of $r_{mn}$ in the $t$-th iteration, and $\tau$ is a constant regularization factor.
{\color{myc1}{According to (\ref{paeq2}) and (\ref{paeq2_1}), we can obtain
\begin{eqnarray}\label{paeq2_2}
&& \!\!\!\!\!\!\!\!\!\!\!\!\!\!\!
r_{sn}\|r_{m_nn}\|_0
  \approx r_{sn}(w_{m_nn}^{(t)} r_{m_nn}+\alpha_{m_nn}^{(t)})
    \nonumber \\
&& \!\!\!\!\!\!\!\!\!\!\!\!\!\!\!
  =\!0.25 w_{m_nn}^{(t)} [(r_{sn}\!+\!r_{m_nn})^2\!-\!(r_{sn}\!- \! r_{m_nn})^2]\!+\!\alpha_{m_nn}^{(t)}r_{sn}
\nonumber \\
&& \!\!\!\!\!\!\!\!\!\!\!\!\!\!\!
\leq0.25 w_{m_nn}^{(t)}(r_{sn}\!+\!r_{m_nn})^2\!-\!0.25w_{m_nn}^{(t)}[(r_{sn}^{(t)}\!- \! r_{m_nn}^{(t)})^2
\nonumber \\
&& \!\!\!\!\!\!\!\!\!\!\!\!\!\!\!\!\quad+2(r_{sn}^{(t)}\!-\!
r_{m_nn}^{(t)})(r_{sn}\! -\!r_{sn}^{(t)}\!- \! r_{m_nn}\! +\! r_{m_nn}^{(t)})]+\alpha_{m_nn}^{(t)}r_{sn}\nonumber \\
&& \!\!\!\!\!\!\!\!\!\!\!\!\!\!\!
\triangleq f_{sm_nn}^{(t)}(r_{sn},r_{m_nn}),
\end{eqnarray}
where the inequality follows from the fact that the quadratic term $(r_{sn}-  r_{m_nn})^2$ is a convex function which is lower bounded by the first order Taylor series.}}

Based on (\ref{paeq2}), (\ref{paeq2_1}) and (\ref{paeq2_2}), the optimization Problem (\ref{max3_1appenB}) after approximation is formulated as:
\begin{subequations}\label{max5_1}
\begin{align}
\mathop{\min}_{ \pmb{r}\geq\pmb 0}& \:\;
\sum_{n=1}^N\sum_{m=1}^{M}
\left(\frac{ \sigma^2}
{H_{mn} }-\frac{ \sigma^2}
{H_{(m+1)n} }\right){2^{\sum_{s=1}^{m}\frac{r_{sn}}{B}}}
 +   \sum_{n=1}^N \sum_{m=1}^M \sum_{s=1}^{m_n}\nonumber\\
&
 \!\! {\color{myc1}{\lambda_m f_{sm_nn}^{(t)}(r_{sn},r_{m_nn})}}
+
\sum_{n=1}^N\frac{\left(\sum\limits_{m=1}^M(w_{mn}^{(t)} r_{mn}+\alpha_{mn}^{(t)})\right)^K}{(L+0.5)^K}
\\
\textrm{s.t.}
&\:\;\sum_{n=1}^N r_{m_n n} = R_m, \quad\forall m.
\end{align}
\end{subequations}

Since the objective function (\ref{max5_1}a) is convex and the constraints (\ref{max5_1}b) are linear, Problem (\ref{max5_1}) is a convex problem, which can be solved by using the popular interior-point method \cite{boyd2004convex}.
We now summarize the proposed joint power control and user clustering (JPCUC) algorithm for solving total power minimization Problem (\ref{max3_1appenB}) in Algorithm 1.
\begin{algorithm}[h]
\caption{JPCUC}
\begin{algorithmic}[1]
\State Initialize a feasible $\pmb r^{(0)}$ of Problem (\ref{max3_1appenB}) and the iteration number $t=0$.
Obtain the values of  $w_{mn}^{(0)}$ and $\alpha_{mn}^{(0)}$ according to (\ref{paeq2}) and
(\ref{paeq2_1}), respectively.
\Repeat:
\State Obtain the optimal $\pmb r^{(t+1)}$ by solving Problem (\ref{max5_1}).
\State Set $t=t+1$, and update the values of $w_{mn}^{(t)}$ and $\alpha_{mn}^{(t)}$ according to (\ref{paeq2}) and
(\ref{paeq2_1}), respectively.
\Until convergence
\end{algorithmic}
\end{algorithm}
%
\subsection{Convergence and Complexity Analysis}
\begin{theorem}
{\color{myc2}{
Starting with any feasible point, the sequence  $\{\pmb r^{(t)}\}_{t=1}^{t=\infty}$  generated by Algorithm 1 is guaranteed to converge to a stationary point of Problem (\ref{max3_1appenB}).}}
\end{theorem}

\itshape \textbf{Proof:}  \upshape
Please refer to Appendix B.
 \hfill $\Box$

For Algorithm 1, the major complexity in each iteration lies in solving the convex optimization (\ref{max5_1}).
Considering that the dimension of the variables in Problem (\ref{max5_1}) is $MN$,
the complexity of solving Problem (\ref{max5_1}) by using the standard
interior point method is $\mathcal O(M^3N^3)$ \cite[Page 487, 569]{boyd2004convex}.
Hence, the total complexity of the Algorithm 1 is $\mathcal O(TM^3N^3)$, where $T$ denotes the total number of iterations.

\section{Numerical Results}
{\color{myc1}{There are $M=10$ users uniformly distributed in a square area of size $300$ m $\times$ $300$ m  with the BS in the center.} }
We set $B=1$ MHz, $\sigma^2=-174$ dBm/Hz, $\alpha_m=0.01$ Joule/Mbits, $\forall m$, and $K=10$.
For propagation model, the large-scale path loss is $L(d)=128.1+37.6\log(d)$ \cite{7437385}, where the unit of $d$ is kilometer, and the standard deviation of shadow fading is $4$ dB.
We consider equal minimum rate demand, i.e., $R_1=\cdots=R_M$.
We compare the proposed JPCUC algorithm for NOMA systems (labeled as `JPCUC-NOMA') with the relax-then-adjust algorithm for NOMA systems in \cite{Lei2016OnPower} (labeled as `RTA-NOMA'),
the suboptimal matching for subcarrier assignment algorithm for NOMA systems in \cite{Fang2016EENOMA} (labeled as `SOMSA-NOMA'),
the dynamic programming algorithm for OMA systems in \cite{Yuan2013Tractability} (labeled as `DP-OMA'),
and the suboptimal matching for subcarrier assignment algorithm \cite{Fang2016EENOMA}  {\color{myc2}{with single-user detection (labeled as `SOMSA-WSD')}}.


According to Fig. 1, NOMA with $L=2$ achieves significant power savings over OMA, especially when the rate demand is high.
This is because the transmission power part dominates the decoding power part which increases with number of users occupying one subcarrier.
{\color{myc1}{It is also observed that the performance of the proposed JPCUC-NOMA with $L=1$  is worse than OMA, since optimal subcarrier assignment is obtained in DP-OMA.}}
Compared with RTA-NOMA, the proposed JPCUC-NOMA consumes lower total power.
{\color{myc1}{This is due to that JPCUC-NOMA does not restrict each subcarrier to be multiplexed by $L=2$ users.}}
From Fig.~2, the JPCUC-NOMA outperforms the other two algorithms for NOMA, which shows the superiority of the proposed algorithm with $\ell_1$ minimization and MM.
{\color{myc1}{The SOMSA-WSD yields the worst performance since SIC is conducted in NOMA to effectively mitigate inter-user interference.
It is shown in Fig.~2 that sum power can be reduced when more users can be multiplexed on each subcarrier, and the energy saving is marginal for large $L$.
This is due to the tradeoff of the transmission power of the BS and the decoding power of the users according to (9).}}

\begin{figure}
\centering
\includegraphics[width=2.6in]{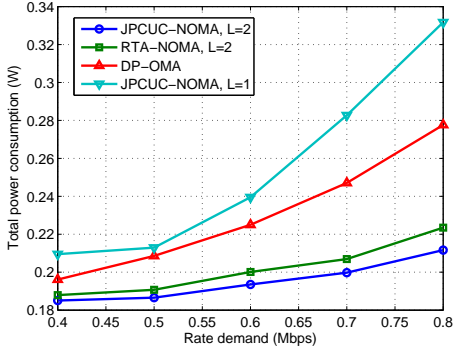}
\vspace{-1em}
\caption{{\color{myc1}{Total power consumption with $N=10$.}} }
\vspace{-1em}
\end{figure}

\begin{figure}
\centering
\includegraphics[width=2.6in]{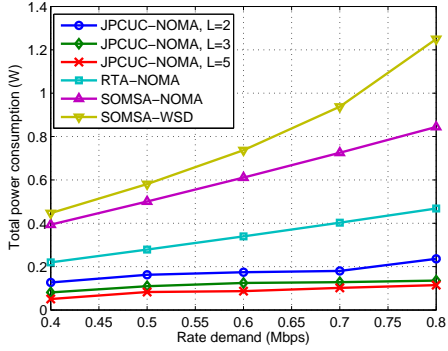}
\vspace{-1em}
\caption{{\color{myc1}{Total power consumption with respect to $L$ with $N=5$. }}}
\vspace{-1.5em}
\end{figure}
\vspace{-0.5em}
\section{Conclusions}
\vspace{-0.25em}
We have formulated the power optimization in NOMA via joint power control and user clustering as an $\ell_0$-norm form.
An efficient algorithm with polynomial complexity is proposed and numerical results show the performance gain of the proposed algorithm with especially the popular matching-based user pairing algorithm.

\appendices
\vspace{-1em}
\section{}
\vspace{-0.25em}
The proof of Theorem 1 is established on showing that  power vector $\pmb p$ can be replaced by rate vector $\pmb r$ without loss of optimality.
According to (\ref{eq2}), we can obtain
\begin{equation}\label{eteq1}
2^{\frac{r_{mn}}{B }} H_{mn} \sum_{l=m+1}^{M} p_{ln}
\! +\!
\sigma^2 \left(2^{\frac{r_{mn}}{B }} -1\right)
\!=\! H_{mn} \sum_{l=m}^{M} p_{ln}, \forall m,n.
\end{equation}
To solve those $MN$ equations, we first define
\begin{equation}\label{eteq2}
a_{mn}=\sum_{l=m}^{M}  p_{ln}, \quad \forall m,n.
\end{equation}
Substituting (\ref{eteq2}) into (\ref{eteq1}) yields
\begin{equation}\label{eteq3}
{a_{mn}}={2^{\frac{r_{mn}}{B }}} a_{{(m+1)n}}+
 \frac {\sigma^2} {H_{mn}}
 \left(2^{\frac{r_{mn}}{B }} -1\right), \quad \forall m,n.
\end{equation}

To solve (\ref{eteq3}), we denote $\pmb a_n=[a_{1n}, \cdots, a_{Mn}]^T$,
\begin{equation}\label{appenAeq2_2_0}\vspace{-0.25em}
\pmb b_n=\left[\frac {\sigma^2} {H_{1n}}
 \left(2^{\frac{r_{1n}}{B }} -1\right), \cdots, \frac {\sigma^2} {H_{Mn}}
 \left(2^{\frac{r_{Mn}}{B }} -1\right)\right]^T,
\end{equation}
and
\vspace{-0.25em}
\begin{equation}\label{appenAeq2_2_2}\vspace{-0.25em}
\pmb W_n=\begin{bmatrix}
0&{2^{\frac{r_{1n}}{B}}}\\
&0 &\!\!\!\!\!\!\!\!{2^{\frac{r_{2n}}{B}}} &  \\
&&\ddots&\ddots\\
&   && 0&{2^{\frac{r_{(M-1)n}}{B }}}\\
&&&& 0
\end{bmatrix}.
\end{equation}
Equations in (\ref{eteq3}) can be rewritten as
\vspace{-0.25em}
\begin{equation}\label{appenAeq2_2_5}\vspace{-0.25em}
(\pmb I - \pmb W_n) \pmb a_n=\pmb b_n,
\end{equation}
where $\pmb I$ is an identity matrix of size $M$.
From (\ref{appenAeq2_2_5}), we have
\begin{equation}\label{appenAeq2_2_6}
\pmb a_n=(\pmb I - \pmb W_n)^{-1}\pmb b_n.
\end{equation}
From the special structure of $\pmb W_n$ in (\ref{appenAeq2_2_2}), we can obtain manipulation $\pmb W_n^l$ for $l\geq1$ by using the recursion method, and specifically $\pmb W_n^{M}=\pmb 0$.
Since
\begin{equation}\label{appenAeq2_2_9}
\vspace{-0.25em}
(\pmb I - \pmb W_n)\left(\pmb I +\sum_{l=1}^{{M-1}} \pmb W_n^l\right)=
\pmb I - \pmb W_n^{M}=\pmb I,
\vspace{-0.25em}
\end{equation}
we have
\begin{equation}\label{appenAeq2_2_9_1}
\vspace{-0.25em}
(\pmb I - \pmb W_n)^{-1}=\pmb I +\sum_{l=1}^{{M-1}} \pmb W_n^l.
\vspace{-0.25em}
\end{equation}

Substituting 
(\ref{appenAeq2_2_9_1}) into (\ref{appenAeq2_2_6}) yields
%
\begin{equation}\label{eteq5}
\vspace{-0.25em}
a_{mn}=\sum_{l=m}^{M}\frac{\sigma^2}{H_{ln}}
\left(2^{\frac{r_{ln}}{B }} -1\right) {2^{\sum_{s=m}^{l-1}\frac{r_{sn}}{B}}}, \quad \forall  m,n,
\vspace{-0.25em}
\end{equation}
where we define ${2^{\sum_{s=m}^{m-1}\frac{r_{sn}}{B}}}=2^0$.
From (\ref{eteq2}) and (\ref{eteq5}), we have
\begin{eqnarray}\label{eteq7_7}
\vspace{-0.25em}
\sum_{m=1}^M p_{mn}&&\!\!\!\!\!\!\!\!\!\!\!=\!a_{1n}\!
=\!\sum_{m=1}^{M}\frac{\sigma^2}{H_{mn}}{2^{\sum_{s=1}^{m}\frac{r_{sn}}{B}}}
-\sum_{m=1}^{M}\frac{\sigma^2}{H_{mn}}{2^{\sum_{s=1}^{m-1}\frac{r_{sn}}{B}}}
\nonumber\\
&&\!\!\!\!\!\!\!\!\!\!\!
=\sum_{m=1}^{M}\left(\frac{\sigma^2}{H_{mn}}-\frac{\sigma^2}{H_{(m+1)n}}\right){2^{\sum_{s=1}^{m}\frac{r_{sn}}{B}}}-
\frac{\sigma^2}{H_{1n}}.
\vspace{-0.25em}
\end{eqnarray}

Equation (\ref{eteq7_7}) implies that power vector $\pmb p$ can be expressed by rate vector $\pmb r$. 
From (\ref{eq2}), we can observe that $p_{mn}=0$ if and only if $r_{mn}=0$, and $p_{mn}>0$ if and only if $r_{mn}>0$.
Thus, we can obtain
\begin{equation}\label{eteq8}
\vspace{-0.5em}
\|\pmb p_n\|_{0}= \|\pmb r_n\|_{0}, \quad \forall n.
\end{equation}
Substituting (\ref{eteq7_7}) and (\ref{eteq8}) into Problem (\ref{max1_1}), we can find that Problem (\ref{max3_1}) is equivalent to Problem (\ref{max1_1}).

\vspace{-1em}
\section{}
\vspace{-0.25em}
%
Since $\ln(1+x)$ is concave, the inequality  $\ln (1+x)^\beta \leq ((1+x_0)^{-1}x+\ln (1+x_0) -(1+x_0)^{-1}x_0)^\beta$ holds for any $x\geq0$, $x_0\geq0$ and $\beta>0$, and achieves equality if and only if $x=x_0$.
Therefore, we can obtain (\ref{max3_1appenB}a) $\leq$ (\ref{max5_1}a)
with equality hold if and only if $r_{mn}=r_{mn}^{(t)}$, $\forall m, n$.

{\color{myc2}{
Algorithm 1 is equivalent to an majorization-minimization (MM) algorithm, which
can be proved to converge 
to a stationary point of the
original problem if the approximate objective function satisfies three conditions.
An MM algorithm is guaranteed to converge to a stationary point of the
original problem if the approximate objective function satisfies the following
conditions according to \cite[Appendix~A]{7437385}:
1) it is continuous,
2) it is a tight upper bound of the original objective function
and
3) it has the same first-order derivative as the original objective function at the point where the upper bound is tight.
Obviously, function in (\ref{max5_1}a) satisfies all these sufficient conditions.
Thus, Algorithm 1 must converge.}}
\vspace{-0.75em}
\bibliographystyle{IEEEtran}
\bibliography{IEEEabrv,MMM}

\end{document}